\documentclass[aps,pre,twocolumn,floatfix,showpacs]{revtex4}
\usepackage{epsf,amsmath,amssymb}
\begin{document}
\title{Disassortativity of random critical branching trees}
\author{J.S. Kim, B. Kahng, and D. Kim}
\affiliation{{Department of Physics and Astronomy, and Center for Theoretical Physics, Seoul National University NS50, Seoul 151-747, Korea}}
\date{\today}
\begin{abstract}
Random critical branching trees (CBTs) are generated by the multiplicative
branching process, where the branching number is determined stochastically,
independent of the degree of their ancestor. Here we show analytically that
despite this stochastic independence, there exists the degree-degree correlation (DDC)
in the CBT and it is disassortative. Moreover, the skeletons of fractal networks,
the maximum spanning trees formed by the edge betweenness centrality, behave
similarly to the CBT in the DDC. This analytic solution and observation
support the argument that the fractal scaling in complex networks originates
from the disassortativity in the DDC.
\end{abstract}
 \maketitle

Recently, it was discovered~\cite{song} that many complex networks in real
world are fractals, satisfying the fractal scaling: The number of
boxes $N_B(\ell)$ needed to cover an object scales in a power-law
manner with respect to the box size $\ell$, i.e., $N_B(\ell)\sim
\ell^{-d_B}$, where $d_B$ is the fractal dimension. Examples are
the World-Wide Web (WWW)~\cite{www}, the protein
interaction network (PIN) of budding yeast~\cite{pin} and the metabolic networks~\cite{metabolic}. In contrast, the Internet~\cite{internet} and many
artificial model networks such as the Barab\'asi-Albert (BA) model~\cite{ba} and
the static model~\cite{static}, etc, are not fractals. It was argued that the
fractal scaling originates from the disassortative correlation
between two neighboring degrees~\cite{yook} or the repulsion between
hubs~\cite{song2}.

The origin of the fractal scaling has been understood from another
perspective~\cite{goh,jskim}: A network is composed of the skeleton,
which is a special type of spanning tree formed
by edges with the highest betweenness centralities or loads, and
the remaining edges in the network that contribute to loop formation.
For fractal networks,
it was shown that the skeletons exhibit fractal
scaling similar to that of the original network. The number
of boxes needed to cover the original network
is almost the same as that needed to cover the skeleton.
Moreover, when a skeleton is considered as a tree generated in a branching
process starting from an arbitrary selected root vertex, the mean branching
number, the average number of offsprings, exhibits a plateau,
albeit fluctuating, independent of the distance from the root. The value
is close to 1, and the skeleton was regarded as the critical
branching tree (CBT), which is known to be a fractal~\cite{burda}. Thus, the
fractal scaling in the original network originates from the presence
of the fractal skeleton underneath the original network.

The CBT is generated by the multiplicative branching process.
To generate a scale-free tree, $n$ ($>0$) offsprings are
created at each branching step with the probability $b_n$, which
is given as follows: $b_n=n^{-\gamma}/\zeta(\gamma-1)$ for $n \ge 1$ and
$b_0=1-\sum_{n=1}^{\infty} b_n$, where $\zeta(x)$ is the
Riemann zeta function. Then the obtained branching tree is a
scale-free network with degree exponent $\gamma$. Since branching event is  stochastically independent, one may think that the CBT is random in the degree-degree correlation (DDC); however, here we show that the DDC is disassortative. We also show that the skeletons
of the fractal networks also exhibit the similar mixing pattern. Therefore, the origin of the
disassortativity of the fractal networks is rooted from the CBT nature of the skeleton.

Here, we calculate the two point correlation function $P(k,k^{\prime})$
for the CBT. $P(k,k^{\prime})$ of an undirected network is defined as
the fraction of links with degrees $k$ and
$k^{\prime}$ on both ends. Even though the network under consideration is
undirected, for further discussion, we make it directed by
assigning arrows to each link in an arbitrary manner.
Then we count the number of links with degree $k$ on the
arrow's source side and $k^{\prime}$ on its
sink side and call it $N(k\to k^{\prime})$. Next reverse all
arrows of the links and count the same and call it $M(k\to
k^{\prime})$. Each link contribute once in $N(k\to k^{\prime})$
and $M(k\to k^{\prime})$. Then
\begin{equation}
P(k, k^{\prime})=\frac{N(k \to k^{\prime})+M(k\to k^{\prime})}{2L},
\end{equation}
where $2L=\langle k \rangle N$ is twice of the link number. Note
that $M(k\to k^{\prime})= N(k^{\prime} \to k)$. This way, a (3-1)
link contributes to the element $P(1,3)$ once and $P(3,1)$ once
while a (2-2) link contribute to $P(2,2)$ twice. Since the sum is
normalized by $2L$, we have the general relation
\begin{equation}
\sum_{k^{\prime}} P(k,
k^{\prime})=\frac{kP_d(k)}{\langle k \rangle}
\end{equation}
with $P_d(k)$ the degree distribution of the network. For uncorrelated networks
only, $P(k,k^{\prime})=kP_d(k)k^{\prime}P_d(k^{\prime})/\langle k
\rangle^2$~\cite{random}.

\begin{figure}[t]
\centerline{\epsfxsize=7.8cm \epsfbox{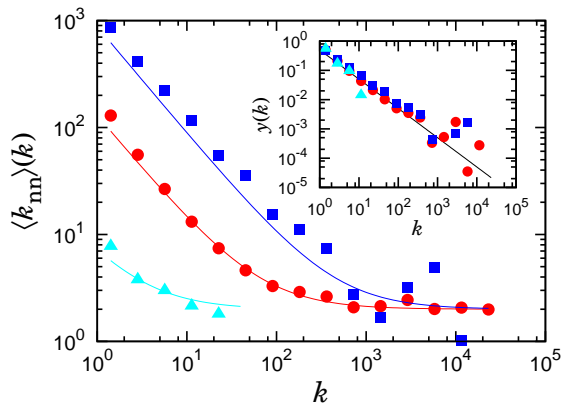}}
\caption{(Color online) Plot of $\langle k_{\rm nn} \rangle (k)$ as a function of degree
$k$ for the CBT ($\circ$) and the skeletons of the fractal networks,
the WWW ($\square$) and the yeast protein interaction network ($\triangle$).
Solid lines represent the formula, Eq.~(\ref{k_nn}).
Inset: To visualize the $1/k$ dependence,
$y(k)=(\langle k_{\rm nn}\rangle (k)-a)/b$ is plotted. Eq.~(\ref{k_nn}) predicts
$y(k)=1/k$, represented by the solid line. The numerical data fit this form reasonably.} \label{fig1}
\end{figure}

Using above procedure, $P(k,k^{\prime})$ for the CBT
with $\langle k \rangle=2$ is obtained as follows:
First, we consider a large enough CBT so that we may neglect the
boundary effect. Assign arrows in the natural
way following the branching direction. Then $N(k\to k^{\prime})$
is the number of degree $k$ nodes ($NP_d(k)$) times the number of
offsprings ($k-1$) times the probability that those offsprings has
degree $k^{\prime}$. So, $N(k\to
k^{\prime})=NP_d(k)(k-1)P_d(k^{\prime})$. Similarly, $M(k\to
k^{\prime})=NP_d(k^{\prime})(k^{\prime}-1)P_d(k)$. So, we find
\begin{equation}
P(k,k^{\prime})=\frac{1}{2}(k+k^{\prime}-2)P_d(k)P_d(k^{\prime}),\label{pkk}
\end{equation}
which is different from the uncorrelated ones, counterintuitively.

The DDC manifests in the mean degree
$\langle k_{\rm nn}\rangle (k)$ of the nearest neighbors of a node
with degree $k$. It is related to $P(k,k^{\prime})$ as
\begin{equation}
\langle k_{\rm
nn}\rangle(k)= \sum_{k^{\prime}}\frac{k^{\prime}
P(k,k^{\prime})}{kP_d(k)/\langle k \rangle},
\label{knn1}
\end{equation}
and is independent of $k$ for uncorrelated networks.
Plugging the formula~(\ref{pkk}) into Eq.~(\ref{knn1}), we obtain that
\begin{equation}
\langle k_{\rm nn}\rangle(k)=\frac{\langle k \rangle^2}{2}+
\frac{\langle k \rangle(\langle k^2 \rangle-2\langle k
\rangle)}{2k}.\label{k_nn}
\end{equation}
$\langle k \rangle=2$ for the CBT.
Eq.~(\ref{k_nn}) may be rewritten in the form, $\langle k_{\rm
nn}\rangle(k)=a+b/k$, where $a=\langle k \rangle^2/2$
and $b=[\langle k \rangle(\langle k^2 \rangle-2\langle k \rangle)]/2$.
Thus, $\langle k_{\rm nn} \rangle(k)$ is inversely proportional
to degree $k$ for $k < b/a$ and thus the CBT is disassortative.

We check this disassortative behavior numerically for the CBT with $\gamma=2.5$
averaged over 10 samples with size $N=10^6$ and show it in Fig.~\ref{fig1}.
Indeed, the numerical data fit well to the
analytic result Eq.~(\ref{k_nn}), represented by the solid line in Fig.~1
and its inset. We note here that for $\gamma<3$, $b/a$ is large
and scales with $N$ as $\sim N^{(3-\gamma)/(\gamma-1)}$, which we
confirm numerically.

\begin{figure}[h]
\centerline{\epsfxsize=7.8cm \epsfbox{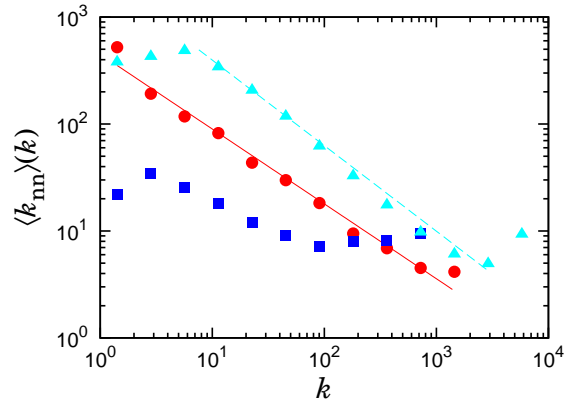}}
\caption{(Color online) Plot of $\langle k_{\rm nn} \rangle (k)$ as a function of degree
$k$ for the skeletons of the non-fractal networks, the Internet at the AS level
($\circ$), the static model with the degree exponent $\gamma=2.4$
($\triangle$), and the BA model with the degree exponent $\gamma=3.0$
($\square$). Solid (dashed) line is a guideline with slope $-0.7$ ($-0.8$).} \label{fig2}
\end{figure}

Next we consider the skeletons of fractal scale-free networks. As discussed above, we argued they could be approximated as CBTs. To corroborate it, we also show in Fig. 1 $\langle k_{\rm nn} \rangle (k)$'s of the skeletons of the WWW and the PIN and
compare them with Eq.~(5) shown as solid lines where the measured values of
$\langle k^2 \rangle$ are used together with $\langle k \rangle=2$.
We find the agreements quite good. Thus, these skeletons can be regarded
as having the same DDC as that of the CBT. On the other hand, the skeletons
of non-fractal networks show different behaviors. Although the Internet at the
autonomous system (AS) level exhibits a disassortative
mixing pattern, the decaying behavior of $\langle k_{\rm nn} \rangle (k)$
for its skeleton is different from that of the CBT as shown in Fig.~\ref{fig2}.
It decays as $\langle k_{\rm nn} \rangle (k)\sim k^{-0.7}.$
For an artificial model, e.g., the static model with degree exponent
$\gamma \approx 2.4 < 3$, $\langle k_{\rm nn} \rangle (k)$ of its skeleton
decays as $\sim k^{-0.8}$ as shown in Fig.~\ref{fig2}, different from
$\sim k^{-1}$ for the CBT. For the BA model with degree exponent $\gamma=3$,
$\langle k_{\rm nn} \rangle (k)$ is random; however, for its skeleton,
it is weakly disassortative for intermediate range of $k$.

In summary, we have shown analytically and numerically that the CBT is disassortative in the DDC. This is induced topologically through the branching process. Its origin is similar to what was proposed~\cite{maslov} and shown later~\cite{juyong} for the Internet that the disassortative mixing pattern is caused by the topological restriction that no pair of nodes is allowed to have multiple connections in the ensemble of graphs with given (or expected) degree sequences. In the CBT, such restriction arises among the offsprings of a  same ancestor that cannot be connected each other. Here, we also showed that the skeletons of fractal complex networks such as the WWW and the PIN exhibit the same pattern in the DDC as found in the CBT. This is yet another evidence that the skeletons of the fractal networks can be regarded as the CBTs besides the mean branching ratio being close to 1
independent of the distance from the root~\cite{goh,jskim}. On the other hand, the skeletons of the non-fractal networks show different patterns in the DDC, even though they are disassortative.
The explicit formula (\ref{pkk}) derived here can be used to study various dynamic problems on fractal networks such as the epidemic problem and so on.

\bigskip\bigskip

This work is supported by KOSEF grant Acceleration Research (CNRC)
(No.R17-2007-073-01001-0) and KRCF.



\begin{thebibliography}{99}
\bibitem{song} C. Song, S. Havlin, and H. A. Makse, Nature (London) {\bf 433},
392 (2005).
\bibitem{www} R. Albert, H. Jeong, and A.-L. Barabasi, Nature (London) {\bf 401},
130 (1999).
\bibitem{pin} We used the dataset by J.-D. Han et al., Nature (London)
{\bf 430}, 88 (2004), which is reported as a fractal in Ref.~\cite{jskim}.
\bibitem{metabolic} H. Jeong, B. Tombor, R. Albert, Z. N. Oltvai, and A.-L.
Barabasi, Nature (London) {\bf 407}, 651 (2000).
\bibitem{internet} University of Oregon Route Views Archive Project, http://
archive.routeviews.org/
\bibitem{ba} A.-L. Barabasi and R. Albert, Science {\bf 286}, 509 (1999).
\bibitem{static} K.-I. Goh, B. Kahng, and D. Kim, Phys. Rev. Lett. {\bf 87}, 278701 (2001).
\bibitem{yook}S.-H. Yook, F. Radicchi, and H. Meyer-Ortmanns, Phys. Rev.
E {\bf 72}, 045105(R) (2005).
\bibitem{song2} C. Song, S. Havlin, and H. A. Makse, Nat. Phys. {\bf 2}, 275
(2006).
\bibitem{goh} K.-I. Goh, G. Salvi, B. Kahng, and D. Kim, Phys. Rev. Lett.
{\bf 96}, 018701 (2006).
\bibitem{jskim} J.S. Kim, K.-I. Goh, G. Salvi, E. Oh, B. Kahng and D. Kim, Phys. Rev.
E {\bf 75,} 016110 (2007).
\bibitem{burda} Z. Burda, J. D. Correia, and A. Krzywicki, Phys. Rev. E {\bf 64},
046118 (2001).
\bibitem{random} S.N. Dorogovtsev, J.F.F. Mendes, A.N. Samukhin, arXiv:cond-mat/0206131.
\bibitem{pastor} R. Pastor-Satorras, A. V\'azquez, and A. Vespignani, Phys. Rev. Lett.
{\bf 87,} 258701 (2001).
\bibitem{maslov} S. Maslov, K. Sneppen, and A. Zaliznyak, Physica A {\bf 333,} 529 (2004).
\bibitem{juyong} J. Park and M.E.J. Newman, Phys. Rev. E {\bf 68,} 026112 (2003).
\end{thebibliography}
\end{document}